\documentclass{elsart}
\input psfig.sty
\usepackage{natbib}
\begin{document}
\runauthor{Eckart, Schinnerer, Tacconi}
\begin{frontmatter}
\title{Molecular Gas and Star Formation in the Host Galaxy of I~Zw~1}

\author[KOL,GAR]{Andreas Eckart}
\author[CAL]{Eva Schinnerer}
\author[GAR]{Linda Tacconi}

\address[KOL]{Universit\"at zu K\"oln, I.Physikalisches Institut,
   Z\"ulpicherstra\ss e 77, 50937 K\"oln, Germany}
\address[GAR]{MPI f\"ur extraterrestrische Physik, 85740 Garching, Germany}
\address[CAL]{Astronomy Department, California Institute of Technology,
   Pasadena, CA 91125, USA}

\begin{abstract}
A recent analysis of high angular resolution NIR imaging and spectroscopic data
in conjunction with Plateau de Bure interferometric
mm-line observations indicate the presence of a circum-nuclear starburst ring
of  $\sim$1.5'' (1.5~kpc) diameter in I~Zw~1.
High angular resolution NIR imaging, using the MPE SHARP camera at the ESO NTT,
HST V-band images, as well as NIR spectroscopy with MPE 3D
provide an improved analysis of the star formation activity
in the disk and nucleus of I~Zw~1.
We present first results from subarcsecond interferometric imaging in the
$^{12}$CO(2-1) line using the Plateau de Bure Interferometer.
\end{abstract}
\begin{keyword}
starbursts, ISM, QSO, I~Zw~1
\end{keyword}
\end{frontmatter}

\section{Introduction}

\begin{figure}[!htp]
\begin{center}
\hspace*{-0.5cm}
\psfig{figure=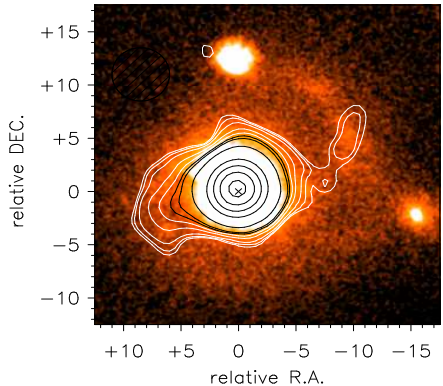,height=6.cm,width=7.0cm,angle=-90.0}
\end{center}
\caption{
\footnotesize
Integrated map of the $^{12}$CO(1-0) line with an angular 
resolution of 5'' corresponding to 5~kpc. 
The contour levels are 0.077, 0.078, 0.080, 0.085, 0.090, 0.095,
0.100, 0.125, 0.150, 0.175, 0.200, 0.250, 0.300, 0.350, 0.400 Jy/beam.
Underlying the contours is an I band HST image of I~Zw~1 convolved
to a resolution of 2'' for comparison.}
\label{fig1}
\end{figure}

The disentangling of the starburst and AGN component in extragalactic objects
and determination of the properties of the interstellar medium and star 
formation in the host galaxies of quasars and QSOs is 
are key problems in the investigation of 
evolutionary sequences and environments of AGNs
(e.g. Genzel et al. 1998, Sanders \& Mirabel 1996,
Norman and Scoville 1988, Sanders et al. 1988, 
Rieke, Lebofsky, and Walker 1988).
At a redshift of z = 0.0611 (Condon, Hutchings, Gower 1985)
I~Zw~1 is the closest QSO for which detailed studies of 
the host galaxy ISM are possible.
All evolutionary sequences involve star formation and an abundant supply of 
material to fuel the starburst and the AGN. 
In order to test these sequences it is 
therefore imperative to investigate the physical properties of the 
molecular gas phase in these objects. For a detailed study of the cold 
molecular interstellar medium, the fluxes and ratios of
the isotopic CO J=2-1 and J=1-0 lines are of interest. 
\\
\\
{\bf General properties of I~Zw~1:}
The optical nucleus of I~Zw~1 is bright with respect to the disk 
(M$_B$=$-$23.45; Schmidt and Green 1983).
This source displays a number of properties found in high redshift QSOs 
(e.g. Buson and Ulrich 1990, Phillips 1978). 
I~Zw~1 is interacting with a nearby companion east of the nuclear bulge
(Fig.\ref{fig1}) and at the edge of its optical disk. The object to the 
north is most likely a foreground star 
(Sargent 1970; Stockton et al. 1982; Eckart et al. 1994).
I~Zw~1 has a 60$\mu$m/100$\mu$m flux density ratio
of 0.9, which is indicative of an even more intense
heating source of UV-visible flux  
than is found in the nearby starburst galaxy M82 (Young and Sanders 1986).

\section{Star Formation and Molecular Gas}

\begin{figure}[!htp]
\begin{center}
\hspace*{0.5cm}
\vspace*{-0.5cm}
\psfig{file=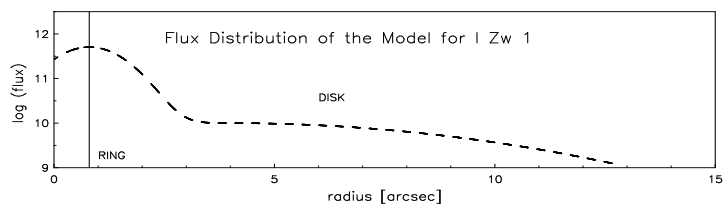,height=3.cm,width=10.0cm,angle=-90.0}
\end{center}
\caption{
\footnotesize
The $^{12}$CO(1-0) line emission as a function of radius for the derived
model. A nuclear ring with a radius of 0.8'' is superposed on an underlying
and much less luminous disk.}
\label{fig2}
\end{figure}

Our {\bf NIR imaging spectroscopy} shows that
the nuclear $K$ band spectrum is dominated by Pa$\alpha$ and Br$\gamma$.
The $H$ band spectrum clearly shows the $^{12}$CO(6-3) band head at 
1.619$\mu$m indicating that about 20\% of the $H$ band continuum is stellar. 
Applying a starburst model shows that about 5\% of the dereddened Br$\gamma$ 
line flux is stellar as well, and that the stellar nucleus 
of the QSO I~Zw~1 shows evidence of a strong decaying starburst.
\\
{\bf $^{12}$CO line emission:}
Combining the stellar mass derived from the starburst model, the dynamical 
mass from the CO-rotation curve and the estimated molecular mass from the 
CO intensity we find that within a factor of two the I$_{CO}$/N$_{H_2}$ 
conversion factor equals the standard value of 
2$\times$10$^{20}$ cm$^{-2}$K$^{-1}$km$^{-1}$s.
The dynamical mass and total molecular gas mass are
3.1$\times$10$^{10}$ M$_{\odot}$
and 5.5$\times$10$^{9}$ M$_{\odot}$, respectively.
\\
As is typical for spiral galaxies, I~Zw~1 exhibits a prominent "double horned" 
$^{12}$CO and HI line profile
(Barvainis, Alloin and Antonucci 1989, Condon, 
Hutchings \& Gower 1985).
Using recent observations of the $^{12}$CO(1-0) line
from the IRAM Plateau de Bure Interferometer (PdBI)
Schinnerer, Eckart \& Tacconi (1998; SET98) were  able to
detect molecular gas 
in the spiral arms of a QSO host galaxy for the first time. 
The 5'' resolution CO-map (Fig.\ref{fig1})
contains the entire single dish $^{12}$CO line flux.
At 2" resolution these interferometric data allowed us to
resolved and separate the 3" to 4" diameter nuclear molecular gas component
from the molecular disk.
The nucleus contains about 2/3 of the total $^{12}$CO(1-0)
line emission.
\\
SET98 have calculated
model position versus velocity diagrams to improve our understanding of the 
structure and dynamics of the nucleus and the disk of I~Zw~1. 
The model projects a given galaxy flux geometry onto a sky grid taking 
account of 
inclination and tilt.  To this flux geometry, a rotation curve is
applied. The beam size and velocity resolution of the observation are
then used to calculate the model p-v diagram.
To model the molecular gas emission in I~Zw~1 ~~SET98 assumed 
a two component Gaussian flux distribution. One Gaussian has
a FWHM of 10'' centered at a radius of 4'' and the other a
FWHM of 1.65'' centered at a radius of 0.8''.
The ratio between the two Gaussian flux peaks is 50, with the
inner one being brighter. This model flux distribution 
is plotted in Fig.\ref{fig2}. 
The azimuthally symmetric modeled p-v diagrams match the
measured ones very well.

\section{New millimeter interferometric data}

\begin{figure}[!htp]
\begin{center}
\hspace*{0.5cm}
\psfig{file=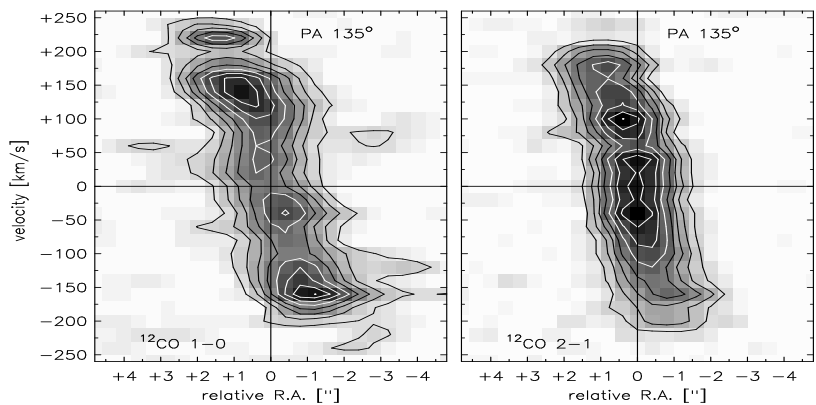,height=7.1cm,width=13.0cm,angle=-90.0}
\end{center}
\caption{
\footnotesize
Position-velocity diagram (grey scale and contours)
of the nuclear ${12}$CO(1-0) (left) and the
$^{12}$CO(2-1) (right) line emission along the major kinematic axis.
Contours are 30\%,40\%, ...100\% of the  peak flux of the 
0.034 Jy/beam (J=1-0 at 1.9'') and the 0.043 Jy/beam
(J=2-1 convolved to 1.9'').}
\label{fig3}
\end{figure}

The most recent  observations of the $^{12}$CO(2-1)
line emission of  I~Zw~1 using the  PdBI
in its B/C configuration have reached subarcsecond angular resolution
(0.9'' FWHM; Schinnerer, Eckart, Tacconi in prep.).
In Fig.\ref{fig3} we compare the  $^{12}$CO(1-0) and  $^{12}$CO(2-1)
line emission p-v diagrams of the nucleus.
The lack of J=1-0 line emission towards the centre reflects the ring-like
distribution of the molecular gas shown in Fig.\ref{fig2}.
Compared to the J=1-0 line, the J=2-1 emission is more centrally peaked.
This is consistent with the finding of Eckart et al. (1994)
that the molecular gas in the disk of I~Zw~1 is cold or sub-thermally excited
(low line ratio I[$^{12}$CO(2-1)]/I[$^{12}$CO(1-0)]$\sim$0.5),
whereas the emission from the nucleus is due to warm optically thick 
molecular gas
(larger line ratio I[$^{12}$CO(2-1)]/I[$^{12}$CO(1-0)]$\sim$1.0)
heated by the nuclear star formation activity.
The increase of the $^{12}$CO(2-1) line intensity due to molecular
excitation, compensates the contrast that is required to get a sharper
view of the ring-like structure seen in the $^{12}$CO(1-0) line emission.
Subarcsecond interferometric $^{12}$CO(1-0) data taken with BIMA are
currently being analyzed to resolve this question
(Staguhn, Schinnerer, Eckart in prep.).

\end{document}